\documentclass[9pt,shortpaper,twoside,web]{ieeecolor}
\usepackage{generic}
\usepackage{cite}
\usepackage{amsmath,amssymb,amsfonts}
\usepackage{algorithmic}
\usepackage{graphicx}
\usepackage{textcomp} 
\usepackage{algorithm,algorithmic}
\usepackage{amsmath}
\usepackage{amssymb}
\usepackage{mathrsfs}
\usepackage{dsfont}
\usepackage{graphicx}
\usepackage{hyperref}
\usepackage{comment}
 
\newtheorem{definition}{\textbf{Definition}}
\newtheorem{theorem}{\textbf{Theorem}}

\newtheorem{lemma}{\textbf{Lemma}}
\newtheorem{remark}{\textbf{Remark}}
\newtheorem{assumption}{\textbf{Assumption}}
\newtheorem{proposition}{\textbf{Proposition}}
\def\BibTeX{{\rm B\kern-.05em{\sc i\kern-.025em b}\kern-.08em
    T\kern-.1667em\lower.7ex\hbox{E}\kern-.125emX}}
\newcommand*{\QEDA}{\hfill\ensuremath{\blacksquare}}
\begin{document}
%\title{A Truthful Profit Maximization Mechanism, Tractable Reformulation, with a Functional Optimization Solution}
%\title{A Truthful Profit Maximization Mechanism Design Using a Tractable Functional Optimization }
\title{A Distributed Primal-Dual Method for Constrained Multi-agent Reinforcement Learning with General Parameterization}
%\title{A Tractable Profit Maximization Mechanism with Autonomous Agent}
%\title{A Tractable Truthful Profit Maximization Mechanism: Autonomous Participation Level Selection}
%\title{A Truthful Tractable Profit Maximization Mechanism: Self Control Agent}
%\title{A Truthful Tractable Profit Maximization Mechanism: Self Allocation Participation Level}
%\author{Mina Montazeri, Hamed Kebriaei,\IEEEmembership{Senior Member,IEEE} and Babak N.Araabi}% <-this % stops a space
%\thanks{Mina Montazeri, Hamed Kebriaei and Babak N.Araabi are with the School of ECE, College of Engineering, University of Tehran, Tehran, Iran. Emails: \{mina.montazeri@ut.ac.ir, kebriaei@ut.ac.ir, araabi@ut.ac.ir\}. }
\author{Ali Kahe, Hamed Kebriaei$^*$, \IEEEmembership{Senior Member, IEEE,}
\thanks{$^*$ Corresponding Author: Hamed Kebriaei.}
\thanks{Ali Kahe and Hamed Kebriaei are with the School of ECE, College of Engineering, University of Tehran, Tehran, Iran. Emails: \{ali.kahe@ut.ac.ir, kebriaei@ut.ac.ir\}.
H. Kebriaei is also with the School of Computer Science, Institute for Research in Fundamental Sciences (IPM), P.O. Box 19395-5746, Tehran, Iran. The work of Hamed Kebriaei was supported in part by the Institute for Research in Fundamental Sciences (IPM) under Grant CS 1404-04-190.}}
%\thanks{$^*$Corresponding author: Hamed Kebriaei.\\
%Mina Montazeri is with the School of ECE, College of Engineering, University of Tehran, Tehran, Iran, and also with the Urban Energy Systems Laboratory, Swiss Federal Laboratories for Materials Science and Technology, Dübendorf, Switzerland. Email: (mina.montazeri@empa.ch).\\
%Hamed Kebriaei and Babak N. Araabi are with the School of ECE, College of Engineering, University of Tehran, Tehran, Iran. Emails: (kebriaei@ut.ac.ir, araabi@ut.ac.ir). }}

%\thanks{Mina Montazeri is with the School of ECE, College of Engineering, University of Tehran, Tehran, Iran. and also with the Urban Energy Systems Laboratory, Swiss Federal Laboratories for Materials Science and Technology, Dübendorf, Switzerland. Emails: \{mina.montazeri@empa.ch\}. }}
%\thanks{The next few paragraphs should contain 
%the authors' current affiliations, including current address and e-mail. For 
%example, F. A. Author is with the National Institute of Standards and 
%Technology, Boulder, CO 80305 USA (e-mail: author@boulder.nist.gov). }
%\thanks{S. B. Author, Jr., was with Rice University, Houston, TX 77005 USA. He is 
%now with the Department of Physics, Colorado State University, Fort Collins, 
%CO 80523 USA (e-mail: author@lamar.colostate.edu).}
%\thanks{T. C. Author is with 
%the Electrical Engineering Department, University of Colorado, Boulder, CO 
%80309 USA, on leave from the National Research Institute for Metals, 
%Tsukuba, Japan (e-mail: author@nrim.go.jp).}}

\maketitle
\begin{abstract}
%Constrained multi-agent reinforcement learning (CMARL) addresses the challenge of optimizing agent's behavior in shared environments while respecting safety, fairness, and resource constraints. 
%\textcolor{red}{del}
This paper proposes a novel distributed approach for solving a cooperative Constrained Multi-agent Reinforcement Learning (CMARL)  problem, where agents seek to minimize a global objective function subject to shared constraints. Unlike existing methods that rely on centralized training or coordination, our approach enables fully decentralized online learning, with each agent maintaining local estimates of both primal and dual variables. Specifically, we develop a distributed primal-dual algorithm based on actor-critic methods, leveraging local information to estimate Lagrangian multipliers. We establish consensus among the Lagrangian multipliers across agents and prove the convergence of our algorithm to an equilibrium point, analyzing the sub-optimality of this equilibrium compared to the exact solution of the unparameterized problem. Furthermore, we introduce a constrained cooperative Cournot game with stochastic dynamics as a test environment to evaluate the algorithm's performance in complex, real-world scenarios. %Experimental results demonstrate the effectiveness of our approach in minimizing the objective cost while keeping constraint violations bounded.

\end{abstract}


\section{Introduction}
Reinforcement Learning (RL) has shown remarkable success in complex decision-making tasks \cite{silver2016mastering,kober2013reinforcement}. While traditional RL focuses on single-agent optimization, many real-world applications like traffic control \cite{Traffic_control},
\begin{comment}
    {finance \cite{finance},}
\end{comment}
and smart grids \cite{smartgrid} require Multi-agent RL (MARL) to handle interacting agents in shared environments \cite{zhang2021multi,nguyen2020deep}. 
A crucial requirement for practical deployment is agent independence during online operation, as centralized approaches often face scalability and communication challenges. This has motivated distributed MARL methods \cite{zhang2023cooperative,cassano2020multiagent,fullydecentralized,yongacoglu2021decentralized}.
% Reinforcement Learning (RL) has gained significant attention in recent years due to its success in solving complex decision-making problems across diverse domains \cite{silver2016mastering}, \cite{kober2013reinforcement}. Traditional RL focuses on minimizing cumulative costs for a single agent or a set of agents acting independently. However, many real-world scenarios—such as traffic control \cite{Traffic_control}, finance \cite{finance}, and smart grids \cite{smartgrid}—involve multiple agents that must cooperate or compete in a shared environment, leading to the study of Multi-agent Reinforcement Learning (MARL). In this framework, agents interact with each other and the environment, seeking to optimize their individual or joint policies while accounting for the influence of other agent's actions (see, e.g., \cite{zhang2021multi}, \cite{nguyen2020deep}, \cite{oroojlooy2023review}). Moreover, a critical requirement for multi-agent systems in real-world scenarios is ensuring the independence of agents during online learning and operation. In many practical applications, centralized controllers face challenges related to scalability, real-time processing, and communication efficiency. To address these issues, distributed methods have been developed \cite{zhang2023cooperative}, \cite{cassano2020multiagent}, \cite{fullydecentralized}, \cite{yongacoglu2021decentralized}.

While MARL holds great promise, many practical applications impose constraints that must be respected to ensure safety, fairness, or efficiency. For example, in networked microgrid management, maintaining power balance and preventing system overloads are critical for system stability \cite{Microgrids}; similarly, in electric vehicle rebalancing systems, agents must consider factors such as battery life and access to charging stations \cite{ElectricVehicle}. Incorporating these types of constraints into reinforcement learning leads to the development of CMARL, which extends MARL frameworks to handle complex environments where satisfying constraints is as crucial as maximizing objective costs.

CMARL is framed within the Constrained Markov Game (CMG) framework \cite{CMG}, where each agent has its own local constraints and objective costs. In this work, we address a cooperative CMARL problem in which all agents aim to minimize a global objective function subject to global constraints, which are composed of local costs. We propose and analyze a distributed online method for solving this cooperative CMARL problem.

Although recent advancements in constrained single-agent reinforcement learning have demonstrated a zero duality gap \cite{zero_duality_Gap}, the duality gap for the cooperative CMARL problem can be non-zero \cite{constrainedMARLHardness}. This inherent complexity poses significant challenges for algorithms attempting to satisfy global constraints \cite{constrainedMARLHardness}. In our study, we provide an analysis on the feasibility and sub-optimality of the equilibrium point of the proposed online algorithm. 

\textbf{Related Works: }
A widely used approach for solving single-agent constrained reinforcement learning problems is the primal-dual method, which demonstrates a zero duality gap by converting the original problem into an unconstrained Markov decision process with a Lagrangian cost \cite{zero_duality_Gap}, \cite{bhatnagar2012online}. This relaxed MDP is solved through alternating updates of the primal and dual variables. This approach has been extended to cooperative CMARL by relaxing the constrained problem into an unconstrained cooperative MARL \cite{diddigi2019actor}, \cite{constrainedMARLHardness}, \cite{descent_ascent_basar}. Such relaxation enables the use of existing MARL algorithms, such as distributed actor-critic methods for networked agents \cite{fullydecentralized}.

Although there are many works on CMARL, only a few address the general problem where agent's local costs are coupled through global constraint functions. For instance, in the CMARL formulation of \cite{First_Order_Constrained_Optimization_2023}, each agent’s actions indirectly influence others through state transition dynamics, but the method requires some coordination, preventing full decentralization. In contrast, \cite{descent_ascent_basar} achieves decentralization through parameter sharing among agents, though it solves a distributed constrained MDP with networked agents rather than a true CMARL problem. This approach assumes homogeneous agents, with policies converging to a consensus. A recent extension by \cite{hassan2024depaint} builds on this by reducing gradient estimation variance for improved scalability.

For general CMARL formulation, \cite{DeCOM} adopts the centralized training, decentralized execution framework, improving computational efficiency and scalability in large-scale multi-agent environments. It separates an agent's policy into two components: a base policy for reward maximization and a perturbation policy for constraint satisfaction. However, it requires communication between agents during execution, distinguishing it from fully distributed methods. 

Another notable work is \cite{javad_lavaei}, which operates in distributed settings. They propose a scalable method for general utility and constraint functions, modeled as nonlinear functions of the state-action occupancy measure. Their approach decomposes the state space for each agent, directly estimating local state-action occupancy measures while leveraging spatial correlation decay and truncated policy gradient estimators for scalability and convergence. However, despite its general policy parameterization, directly estimating local occupancy measures remains challenging in large state-action spaces.

\textbf{Contributions: }In this work, we propose a distributed algorithm for the CMARL problem with networked agents, where each agent maintains its own local primal and dual variables. Specifically, we estimate the Lagrangian multipliers of the global CMARL problem using only local information, enabling a fully distributed policy update method. To the best of our knowledge, this problem has not been addressed previously. Our approach builds on the method of 
%\cite{bhatnagar2012online} for constrained single-agent RL and 
\cite{fullydecentralized} for unconstrained MARL. As such, our method can be viewed as a 
%multi-agent extension of \cite{bhatnagar2012online} and a
constrained extension of \cite{fullydecentralized}. The main contributions of the article are as follows.
{
\begin{enumerate}
    \item  We propose a distributed formulation for Lagrange multipliers in the cooperative CMARL problem over networked agents.
    \item We develop a distributed online primal-dual algorithm based on actor-critic methods with general function approximation.
    \item We prove the consensus of the Lagrange multipliers and the convergence of the proposed algorithm to an equilibrium point.
    \item  We analyze feasibility of constraints, and sub-optimality of the algorithm’s equilibrium point compared to the exact solution of the unparameterized primal CMARL problem.
\end{enumerate}}

\textbf{Notation: } 
Let $\mathbb{R}^d$ denote the $d$-dimensional Euclidean space, with $\|\cdot\|$ representing the Euclidean (vector) and spectral (matrix) norms, and $\|\cdot\|_1$ the $L_1$ norm. Let $\nabla$ denote the gradient operator, $\mathbf{1}$ the all-ones vector, $\mathds{1}_A$ the indicator of set $A$, and $\mathbf{I}_d$ the $d \times d$ identity matrix. 
Define the consensus projection matrix as $J := \frac{1}{N}\mathbf{1}\mathbf{1}^\top \otimes \mathbf{I}_d$, where $\otimes$ denotes the Kronecker product, and its orthogonal complement as $J_\perp := \mathbf{I}_{Nd} - J$.
%Let $\mathbb{R}^d$ be $d$-dimensional Euclidean space with $\|\cdot\|$ denoting the Euclidean (vector) and spectral (matrix) norms. Define $\|\cdot\|_1$ as the $L_1$ norm, $\nabla$ as the gradient operator, $\mathbf{1}$ as the all-ones vector, $\mathds{1}_{A}$ as the indicator function for set $A$, and $\mathbf{I}_d$ as the $d\times d$ identity matrix. The consensus projection matrix is $J := \frac{1}{N}\mathbf{1}\mathbf{1}^\top \otimes \mathbf{I}_d$ (where $\otimes$ is Kronecker product), with its orthogonal complement $J_\perp := \mathbf{I}_{Nd} - J$.
%Let \(\mathbb{R}^d\) denote the \(d\)-dimensional Euclidean space, and let \(\|\cdot\|\) represent the Euclidean norm for vectors and the spectral norm for matrices. The \(L_1\) norm, denoted by \(\|\cdot\|_1\), is the sum of the absolute values of the vector components. We define the gradient operator by \(\nabla\). The notation \(\mathbf{1}\) represents a vector of ones, and \(\mathds{1}_{A}\) denotes the indicator function for the set \(A\). Additionally, \(\mathbf{I}_d\) denotes the identity matrix of size \(d \times d\). We define the matrix $J$ as the projection matrix onto the consensus subspace, given by $J:= \frac{1}{N} \mathbf{1}\mathbf{1}^\top \otimes \mathbf{I}_d,$ where $\otimes$ denotes the Kronecker product. The orthogonal projection onto the subspace orthogonal to the consensus subspace is given by: $J_\perp:= \mathbf{I}_{Nd} - J.$
For any vector $x \in \mathbb{R}^{Nd}$, the consensus component is defined by $\langle x \rangle := \frac{1}{N}(\mathbf{1}^\top \otimes \mathbf{I}_d)x=\frac{1}{N}(x_1 + \dots + x_N)$, where \(x_i \in \mathbb{R}^d\) for $\{i = 1, \dots, N\}$.  
The disagreement component of \(x\) relative to the consensus subspace is given by \(x_\perp := J_\perp x\). Thus, any vector \(x\) can be decomposed as $x = \mathbf{1} \otimes \langle x \rangle + x_\perp.$ Furthermore, we denote the projected dynamical system corresponding to the projection operator \( {P}_C: \mathbb{R}^m \rightarrow C \)  using $\dot{x} = \hat{{P}}_C[f(x)],$ where \(\hat{{P}}_C[f(x)] = \lim_{\varepsilon \rightarrow 0^+} \left[\frac{{P}_C[x + \varepsilon f(x)] - x}{\varepsilon}\right]\). {$\mathbb{R}_{+}$ and $\mathbb{R}_{++}$ denote the set of non-negative and strictly positive real numbers, respectively.}

\section{Problem Formulation of Distributed CMARL}
%In this section, 
We formally define the cooperative CMARL problem based on the CMG framework %We then propose a novel distributed problem formulation for CMARL, in which networked agents update their policies using only local information.
%To begin,
%We consider a CMG
which is represented as a tuple \( \text{CMG} = \{\mathcal{S}, \mathcal{A}, P, \mathcal{C}, \mathcal{G}, b\} \). This framework involves \(N\) agents, subject to \(K\) constraints, where agents and constraints are indexed by \(n \in \mathcal{N} = \{1, \cdots, N\}\) and \(k \in \mathcal{K} = \{1, \cdots, K\}\), respectively. The components of the CMG tuple include the following: The set \( \mathcal{S} \) is the finite state space. At each time step \( t \), a random state \( s^t \in \mathcal{S} \) is drawn. 
{
% R2.C10 
The joint action space is denoted by  \( \mathcal{A} = \prod_{n=1}^{N} \mathcal{A}_n \), where \( \mathcal{A}_n \) is the finite action space of agent \( n \). The joint action of all agents is denoted by \( a = (a_1, a_2, \ldots, a_N)^\top \), where \( a_n \) is the action of agent \( n \). The vector \( a_{-n} = (a_1, \ldots, a_{n-1}, a_{n+1}, \ldots, a_N)^\top \) represents the joint actions of all agents except agent \( n \), with \( a_{-n} \in \mathcal{A}_{-n} = \prod_{j \neq n} \mathcal{A}_j \).} Thus, the joint action can be succinctly expressed as \( a = (a_n, a_{-n}) \). \( P(s'|s, a) \) is the probability transition kernel, representing the probability of transitioning from state \( s \in \mathcal{S} \) to state \( s' \in \mathcal{S} \) under the joint action \( a \in \mathcal{A} \). {
% R2.C10 
 The set \( \mathcal{C} = \{c_n(s, a) \mid n \in \mathcal{N}\} \) represents the local expected cost functions for each agent. Each function \( c_n(s, a): \mathcal{S} \times \mathcal{A} \rightarrow \mathbb{R} \) defines the expected cost incurred by agent \( n \) at state \( s \) under the joint action \( a \).} Specifically, \( c_n(s, a) = \mathbb{E}[c_n^{t+1} \mid s^t = s, a^t = a] \), where \( c_n^{t+1} \) is the immediate cost incurred by agent \( n \) at time step \( t+1 \). {
The set $\mathcal{G} = \{g_{n,k}(s,a) \mid n \in \mathcal{N}, k \in \mathcal{K}\}$ denotes the local expected constraint cost functions. Each $g_{n,k}: \mathcal{S} \times \mathcal{A} \rightarrow \mathbb{R}$ specifies the expected $k$-th constraint cost incurred by agent $n$ at state $s$ and joint action $a$, i.e., $g_{n,k}(s,a) = \mathbb{E}[g_{n,k}^{t+1} \mid s^t = s, a^t = a]$, where $g_{n,k}^{t+1}$ is the immediate cost at time $t+1$. This formulation allows exogenous randomness in objective and constraint costs at each state--action pair. The vector $\mathbf{b} = (b_1, \ldots, b_K)^\top$ defines the upper bounds, where $b_k$ limits the $k$-th constraint cost.

 % R2.C10 

%The set \( \mathcal{G} = \{g_{n,k}(s, a) \mid n \in \mathcal{N}, k \in \mathcal{K}\} \) represents the local expected constraint cost functions for each agent. Each function \( g_{n,k}(s, a): \mathcal{S} \times \mathcal{A} \rightarrow \mathbb{R} \) defines the expected cost for the \( k \)-th constraint incurred by agent  \( n \) under state \( s \) and joint action \( a \), i.e., \( g_{n,k}(s, a) = \mathbb{E}[g_{n,k}^{t+1} \mid s^t = s, a^t = a] \), where \( g_{n,k}^{t+1} \) is the immediate cost for \( k \)-th constraint incurred by agent \( n \) at time step \( t+1 \). It is worth noting that in this formulation, we have considered exogenous sources of randomness for objective and constraint costs at each state-action pair. The vector \( \mathbf{b} = (b_1, b_2, \ldots, b_K)^\top \) defines the upper bounds for the constraint functions, where each \( b_k \) specifies the limit for the \( k \)-th constraint cost.}

CMARL is formulated within the CMG framework, where each agent $n$ adopts an individual policy $\pi_n \in \mathcal{P}_n := \Delta(\mathcal{A}_n)^{\mathcal{S}}$, with $\Delta(\mathcal{A}_n)$ denoting the probability simplex over $\mathcal{A}_n$. As agents act independently, the joint policy factorizes as $\pi = \prod_{n \in \mathcal{N}} \pi_n \in \mathcal{P} = \prod_{n \in \mathcal{N}} \mathcal{P}_n$.
We consider parameterized policies, where each agent $n$ has parameters $\theta_n \in \Theta_n \subseteq \mathbb{R}^m$, with $\Theta_n$ compact and convex, enabling efficient representation in large state spaces. The global parameter vector is $\theta = [\theta_1^\top, \cdots, \theta_N^\top]^\top \in \Theta$, where $\Theta = \prod_{n \in \mathcal{N}} \Theta_n$, and the corresponding product policy is $\pi_\theta(a|s) = \prod_{n \in \mathcal{N}} \pi_{\theta_n}(a_n|s)$.
Define $\theta_{-n} = [\theta_1^\top, \cdots, \theta_{n-1}^\top, \theta_{n+1}^\top, \cdots, \theta_N^\top]^\top \in \Theta_{-n}$, where $\Theta_{-n} = \prod_{j \neq n} \Theta_j$. Thus, $\theta = (\theta_n, \theta_{-n})$.

\begin{assumption}
\label{non_zero policy}
    For any agent \( n \in \mathcal{N} \), state \( s \in \mathcal{S} \), and action \( a_n \in \mathcal{A}_n \), the policy \( \pi_{\theta_n}(a_n | s) > 0 \) holds for all \( \theta_n \in \Theta_n \). Moreover, the policy \( \pi_{\theta_n}(a_n | s) \) is continuously twice differentiable with respect to \( \theta_n \).
\end{assumption}

\begin{assumption}
\label{ergodicity}
    Under any product parameterized policy, the Markov chain \( \{s^t\}_{t=0}^{\infty} \) is assumed to be irreducible and aperiodic.
\end{assumption}
{
%R2.C5

According to the previous assumptions, the Markov chain induced by any product parameterized policy $\pi_\theta$ admits a unique stationary distribution, denoted by $d^{\pi_\theta}(s)$.} 
Additionally, the class of parameterized softmax policies is a rich policy class that can induce Markov chains satisfying Assumption \ref{ergodicity} 
under appropriate transition probability functions.

\begin{assumption}
\label{signal boundedness}
The sequence \( \{|c_n^{t+1}|, |g_{n,k}^{t+1}| \mid \forall n \in \mathcal{N}, \forall k \in \mathcal{K}\}_{t \geq 0} \) is real-valued, uniformly bounded, and { its elements mutually are conditionally independent given the state-action pairs} at each time step \( t \geq 0 \).
\end{assumption}
{
%R2.C5 

This assumption allows the cost samples for each agent to be treated as bounded and i.i.d., which is standard in statistics settings.
}
\begin{remark}
	Without loss of generality, we assume that both the immediate and constraint costs are bounded in $[0,1]$.
    %Without loss of generality, we assume that both the immediate cost and the constraint cost are bounded within the interval \( [0, 1] \).
\end{remark}
In CMARL, agents cooperatively seek an optimal feasible policy that minimizes the \textit{global objective cost}, defined as the average of local objective costs across all agents. Although each agent acts independently using only local information, they jointly aim to minimize this global objective. The global objective cost $J(\pi_{\theta})$ is formally defined as:
%	In CMARL, agents collaborate to find an optimal, feasible policy that minimizes the \textit{global objective cost function}, which is defined by the total average of the local objective costs across all agents. Although each agent operates independently and utilizes only its local information, their collective goal is to minimize this global objective function. The formal definition of the global objective cost \( J(\pi_{\theta}) \) is expressed as:
\begin{align}
    \label{global objective cost}
    J(\pi_{\theta}) & = \lim_{T \to \infty} \frac{1}{T} \mathbb{E}^{\pi_{\theta}} \left[ \sum_{t=0}^{T{-1}} \left( \frac{1}{N} \sum_{n \in \mathcal{N}} c_{n}^{t+1} \right) \right]  \\ 
    & = \sum_{s \in \mathcal{S}} d^{\pi_{\theta}}(s) \sum_{a \in \mathcal{A}} \pi_\theta(a|s) c(s,a) \nonumber
\end{align}
Here, the local cost function is defined as \( c(s,a) = \frac{1}{N}\sum_{n \in \mathcal{N}} c_n(s,a) \), representing the average cost across all agents. In a similar manner to the global objective cost, we define the global constraint functions, which agents are expected not to violate. The global constraint functions are given by:
\begin{align}
    \label{constraints function}
    G_k(\pi_{\theta}) &= \lim_{T \to \infty} \frac{1}{T} \mathbb{E}^{\pi_{\theta}} \left[ \sum_{t=0}^{T{-1}} \left( \frac{1}{N} \sum_{n \in \mathcal{N}} g_{n,k}^{t+1} \right) \right] \\
    &= \sum_{s \in \mathcal{S}} d^{\pi_{\theta}}(s) \sum_{a \in \mathcal{A}} \pi_\theta(a | s) g_k(s,a) \leq b_k, \quad \forall k \in \mathcal{K} \nonumber
\end{align}
where \( g_k(s, a) = \frac{1}{N} \sum_{n \in \mathcal{N}} g_{n,k}(s, a) \) represents the average constraint cost across all agents for the \( k \)-th constraint. Additionally, we denote \( G(\pi_\theta) = (G_1(\pi_{\theta}), \cdots, G_K(\pi_{\theta}))^\top \).{ The second equality in \eqref{global objective cost}  and \eqref{constraints function} holds true under Assumptions \ref{non_zero policy}-\ref{signal boundedness} \cite[Proof of (9.5)]{zhao2025mathematical}.} It is important to note that the global constraint functions must be satisfied collectively by all agents and are not specific to any single agent. In fact, the constraint functions in \eqref{constraints function} impose implicit restrictions on the product parameter space \( \Theta \). Therefore, we define the feasible product parameter space as \( \Theta_C \), where $\Theta_C := \{\theta \in \Theta \mid G_k(\pi_{\theta}) \leq b_k, \, \forall k \in \mathcal{K}\}$.
The global Lagrangian function for this stochastic multi-agent optimization problem is then expressed as:
\begin{align}
    \label{global lagrangian}
    \mathcal{L}(\pi_{\theta}, \lambda) &= J(\pi_{\theta}) + \sum_{k=1}^{K} \lambda_k \left( G_k(\pi_{\theta}) - b_k \right)\\
    &= \sum_{s \in \mathcal{S}} d^{\pi_{\theta}}(s) \sum_{a \in \mathcal{A}} \pi_{\theta}(a | s) L(s,a;\lambda) \nonumber
\end{align}
where \( L(s, a; \lambda) = c(s, a) + \sum_{k=1}^{K} \lambda_k \left( g_k(s, a) - b_k \right) \) represents the expected global Lagrangian cost, \(\lambda = (\lambda_1, \ldots, \lambda_K)^\top\) is the global Lagrange multiplier vector associated with the global constraint functions, and \(\lambda_k \geq 0\) for all \(k \in \mathcal{K}\). { Now, by introducing $L_n(s,a;\lambda)=c_n(s,a)+\sum_{k=1}^{K}\lambda_k g_{n,k}(s,a)$ we can rewrite \eqref{global lagrangian} as follows: 
\begin{align}
  \mathcal{L}(\pi_{\theta}, \lambda) = \sum_{s \in \mathcal{S}} d^{\pi_{\theta}}(s) \sum_{a \in \mathcal{A}} \pi_{\theta}(a | s) \left( \frac{1}{N} \sum_{n \in \mathcal{N}} L_n(s, a; {\lambda}) \right) \
\end{align}
Consequently, the global Lagrangian function can be expressed as the summation of local Lagrangian functions as shown below.
\begin{align}
 \label{decomposable}
 \mathcal{L}(\pi_{\theta}, \lambda) = 	\frac{1}{N}\sum_{n \in \mathcal{N}} 	\mathcal{L}_n(\pi_{\theta}, \lambda)
\end{align}
where $\mathcal{L}_n(\pi_{\theta}, \lambda)= \sum_{s \in \mathcal{S}} d^{\pi_{\theta}}(s) \sum_{a \in \mathcal{A}} \pi_{\theta}(a | s) L_n(s, a; {\lambda}) $.}
%It should be mentioned that although $\mathcal{L}(\pi_{\theta}, \lambda)$ in \eqref{decomposable} has a decomposed form, all agents should have access to the Lagrange multipliers, which are coupling variables. Therefore, addressing this optimization problem is not possible by adopting distributed approaches in this setting.}

Under this formulation, the cooperative CMARL problem is cast as the minimax problem 
$\inf_{\theta \in \Theta} \sup_{\lambda \in \mathbb{R}_{+}^K} \mathcal{L}(\pi_{\theta}, \lambda)$, with dual 
$\sup_{\lambda \in \mathbb{R}_{+}^K} \inf_{\theta \in \Theta} \mathcal{L}(\pi_{\theta}, \lambda)$. 
Due to a nonzero duality gap~\cite{constrainedMARLHardness}, the primal and dual may differ. 
We therefore solve the dual, which is better suited to distributed constrained optimization~\cite{chang2014distributed,nedic2009distributed}. 
For fixed multipliers, the inner minimization reduces to an unconstrained MARL problem~\cite{fullydecentralized}, where agents cooperatively minimize the global Lagrangian using local information.
Our algorithm alternates between policy updates and multiplier adjustments. Although $\mathcal{L}(\pi_{\theta}, \lambda)$ in~\eqref{decomposable} is decomposed, the shared Lagrange multipliers remain coupling variables, preventing fully distributed solutions.
To enable distributed CMARL, we introduce \textit{locally estimated Lagrange multipliers} $\{\hat{\lambda}_{n,k} \mid n \in \mathcal{N}, k \in \mathcal{K}\}$, where $\hat{\lambda}_{n,k}$ is agent $n$’s estimate of the global multiplier $\lambda_k$. The local estimate vector for agent $n$ is $\hat{\lambda}_n = (\hat{\lambda}_{n,1}, \cdots, \hat{\lambda}_{n,K})^\top$, and the joint vector across all agents is $\hat{\lambda} = (\hat{\lambda}_1^\top, \cdots, \hat{\lambda}_N^\top)^\top$.

% considering that each agent operates with local parameters \( (\theta_n, \hat{\lambda}_n) \) , and aim to do this process in a distributed manner, using only local information

    \begin{definition}[Estimated Lagrangian Objective]
        Let $\{\hat{\lambda}_n\}_{n \in \mathcal{N}}$ denote local estimates of the Lagrange multipliers.  
        The \emph{estimated Lagrangian objective function} associated with policy $\pi_\theta$ is defined as:
        \begin{align}
            \label{distributed lagrangian function}
            \hat{\mathcal{L}}(\pi_{\theta}, \hat{\lambda}) &:= \sum_{s \in \mathcal{S}} d^{\pi_{\theta}}(s) \sum_{a \in \mathcal{A}} \pi_{\theta}(a|s) \left( \frac{1}{N} \sum_{n \in \mathcal{N}} L_n(s, a; \hat{\lambda}_n) \right)  \nonumber \\
            &= \frac{1}{N}\sum_{n \in \mathcal{N}} \mathcal{L}_n(\pi_{\theta}, \hat\lambda_n)
        \end{align}
        \end{definition}

{Where \( L_n(s, a; \hat{\lambda}_n) = c_n(s, a) + \sum_{k=1}^{K} \hat{\lambda}_{n,k} \left( g_{n,k}(s, a) - b_k \right) \) represents the expectation of estimated local Lagrangian cost with respect to  $\hat{\lambda}$ for agent \( n \), \(\mathbb{E}[l_n^{t+1}(\hat{\lambda}_n) \mid s^t = s, a^t = a] \). Accordingly, we can define the immediate local Lagrangian cost for agent \( n \) at time step \( t+1 \) as \(l_n^{t+1}(\hat{\lambda}_n) := c_{n}^{t+1} +\Sigma_{k=1}^K \hat{\lambda}_{n,k} (g_{n,k}^{t+1}-b_k) \), and 
the immediate global Lagrangian cost at time step \( t+1 \), aggregated over all agents, as \( l^{t+1}(\hat{\lambda}) = \frac{1}{N} \sum_{n \in \mathcal{N}} l_n^{t+1}(\hat{\lambda}_n)\).}

For given locally estimated multipliers, $\hat{\mathcal{L}}(\pi_{\theta}, \hat{\lambda})$ serves as an unconstrained objective, solvable via distributed actor--critic methods~\cite{fullydecentralized}. Moreover, if the estimates satisfy the consensus condition $\hat{\lambda}_{n,k} = \lambda_k$ for all $n \in \mathcal{N}$ and $k \in \mathcal{K}$, then the estimated Lagrangian coincides with the global Lagrangian in~\eqref{global lagrangian}, i.e., $\hat{\mathcal{L}}(\pi_{\theta}, \hat{\lambda}) = \mathcal{L}(\pi_{\theta}, \lambda)$.
%Therefore For a given set of locally estimated Lagrange multipliers, $\hat{\mathcal{L}}(\pi_{\theta}, \hat{\lambda})$ can be treated as an unconstrained objective cost function, which is solvable using distributed actor-critic methods, as described in \cite{fullydecentralized}. Furthermore, in this new setting, when the locally estimated Lagrange multipliers satisfy the consensus condition \( \hat{\lambda}_{n, k} = \lambda_k \) for all \( n \in \mathcal{N} \) and \( k \in \mathcal{K} \), the estimated Lagrangian objective function becomes equivalent to the global Lagrangian function in \eqref{global lagrangian}, i.e., \( \hat{\mathcal{L}}(\pi_{\theta}, \hat{\lambda}) = \mathcal{L}(\pi_{\theta}, \lambda)\). 
%We will discuss how this consensus condition is satisfied for our proposed algorithm in details in Theorem \ref{lambda_perp}. Also, it can be considered that each agent can operates independently with its local parameters \((\theta_n, \hat{\lambda}_n)\) in the environment, using this setting. So, this formulation allows us to design distributed algorithms with local information for each agent.
Theorem \ref{lambda_perp} will provide detailed convergence guarantees for this consensus condition. %Notably, this formulation enables distributed implementation, where each agent operates independently using only local parameters $(\theta_n, \hat{\lambda}_n)$ and environmental interactions.

Accordingly, for a continuously differentiable policy \( \pi_{\theta} \) and given \( \hat{\lambda} \), the gradient of \(\hat{\mathcal{L}}(\pi_\theta,\hat{\lambda}) \) with respect to the parameters \( \theta_n \) is given by the policy gradient theorem for MARL \cite{fullydecentralized}:
\begin{align}
    \label{policy gradient}
    \nabla_{\theta_n} \hat{\mathcal{L}}(\pi_{\theta}, \hat{\lambda}) = \mathbb{E}^{\pi_\theta} \left[ \nabla_{\theta_n} \log \pi_{\theta_n} A_n^{\pi_{\theta}}(s, a) \right],
\end{align}
where \( A_n^{\pi_{\theta}}(s, a) \) is the advantage function for agent \( n \), defined as:
\begin{equation}
A_n^{\pi_{\theta}}(s, a)=Q^{\pi_{\theta}}(s, a) - \sum_{a_n \in \mathcal{A}_n} \pi_{\theta_n}(a_n | s) Q^{\pi_{\theta}}(s, (a_n, a_{-n})). \nonumber
\end{equation} 
The term \( Q^{\pi_{\theta}}(s, a) \) refers to the global differential expected action-value function of $\hat{\mathcal{L}}(\pi_{\theta}, \hat{\lambda})$, and is defined as follows: % \cite{sutton2018reinforcement}:
$$Q^{\pi_{\theta}}(s, a) = \sum_{t=0}^{\infty} \mathbb{E}^{\pi_{\theta}} \left[ l^{t+1}(\hat{\lambda}) - \hat{\mathcal{L}}(\pi_{\theta}, \hat{\lambda}) \mid s^t = s, a^t = a \right].$$
{In this work, we linearly parameterize the state-action value function approximator \( Q(s, a) \) as $Q(s, a; w) = w^\top\phi(s, a),$ where \( \phi(s, a) \) denotes the feature vector associated with the state-action pair. Also, we define $\phi^t:=\phi(s^t, a^t)$ for simplicity in notation. So, in the multi-agent framework, each agent $n$ maintains its own local weight vector \( w_n \in \mathbb{R} \) , which is used to approximate its individual value function.}
Indeed, the approximation of the Q-function may introduce an error relative to the true policy gradient in \eqref{policy gradient}.

\section{Algorithm}
 To solve the presented CMARL problem,  a distributed primal-dual algorithm is proposed which works based on the actor-critic method. 
	The actor updates parameter $\theta_n$  of parameterized stochastic policy $\pi_{\theta_n}(\cdot\mid s)$ that maps states to a distribution over actions to improve long-term performance using policy-gradient (7). 
	The critic approximates the expected cumulative cost incurred when action $a$ is taken in state $s$ and the current policy is followed thereafter, This is done by updating  parameters $w_n$  of local action-value function $Q(s,a;w_n)$.
	The critic parameters $w_n$ are updated using \emph{temporal-difference (TD) learning} by minimizing the discrepancy between the current value estimate and a one-step target that combines the observed immediate cost with the estimated value of the subsequent state--action pair. 
%	This discrepancy is captured by the TD error, which drives the critic update at each iteration.	
%	These actor, critic, and TD-learning components are combined with decentralized consensus and primal--dual updates to address the constrained multi-agent reinforcement-learning problem considered in this work.
	\\
	The proposed algorithm is outlined in Algorithm \ref{alg:alg1}.
	All agents initialize their policy parameters, critic weights, constraint estimates, and Lagrange multipliers (line 1). An initial joint action is sampled from the local policies (line 2), and the iteration counter is set (line 3). Then we have the following two phases:\\
	\textit{Phase I — Sampling and Lagrangian cost update:}
	At time step $t$, each agent observes the next state $s^{t+1}$, along with its local scalar objective cost $c_n^{t+1}$ and vector-valued constraint cost $g_n^{t+1}$.  (line 6). These values are combined with the current Lagrange multiplier estimate to form an immediate local Lagrangian cost $l_n^{t+1}$ (line 7). This immediate cost is used to update the value of locally estimated Lagrangian $\hat{\mathcal{L}}_n^{t+1}$ (line 8). The agent then samples and executes its next action using the current policy (line 9). 
	After all, each agent can observe the next joint action $a^{t+1}=(a^{t+1}_n,a^{t+1}_{-n})$, which includes both its action and the actions of the other agents (line 11).\\
	\textit{Phase II — Learning parameters and consensus:}
 In this phase, each agent updates its critic, actor, and dual variable parameters using local information to solve a decomposed\ multi-agent Lagrangian objective. %To do so, each agent computes a surrogate immediate cost $l_n^{t+1}(\hat{\lambda}_n)$, which integrates the objective and constraint information into a single objective signal. This phase builds upon Algorithm~1 from \cite{fullydecentralized}. 
 During this step, each agent is required to share its local critic parameters with neighbors defined by the communication graph. The associated weight matrix of this communication graph at time step $t$ is denoted by $\Xi^t = [\xi_{n,n'}^t]$ for all $n, n' \in \mathcal{N}$. The critic update rule for each agent $n$ in this phase is given by (lines 14 and 19):

\begin{align}
    \label{critic update}
    &\hat{\mathcal{L}}_{n}^{t+1} = \hat{\mathcal{L}}_{n}^{t} + \alpha^t \left( l_n^{t+1}(\hat{\lambda}_n^t) - \hat{\mathcal{L}}_{n}^{t} \right), \nonumber \\ 
    &\Tilde{w}_n^t = w_n^t  + \alpha^{t} \delta_n^{t} \nabla_w Q_{{\lambda}}(s^{t}, a^{t}; w_n^t), \\
    &w_n^{t+1} = \sum_{n' \in \mathcal{N}} {\xi}_{n,n'}^{t+1} \Tilde{w}_{n'}^t. \nonumber
\end{align}
Here, $\alpha^t$ represents the critic learning rate, and the temporal difference (TD) error is updated using the following rule (line 13):
\begin{equation}
    \delta_n^{t} = l_n^{t+1}(\hat{\lambda}_n^t) - \hat{\mathcal{L}}_{n} + Q_{{\lambda}}(s^{t+1}, a^{t+1}; w_n^t) - Q_{{\lambda}}(s^t, a^t; w_n^t).
\end{equation}
Then, based on the policy gradient theorem \eqref{policy gradient}, each agent updates its actor parameter $\theta_n$ locally using sample-based gradient estimation of Lagrangian objective cost as follows (lines 16-18):
\begin{align}
    \label{actor_update}
    A_n^t = Q_{{\lambda}}(s^{t}, a^{t}; w_n^t) - &\sum_{a_n \in \mathcal{A}_n} \pi_{\theta_n^t}(a_n | s^{t+1}) Q_{{\lambda}}(s^t, (a_n^t, a_{-n}^t); w_n^t), \nonumber \\
    &\psi_n^t = \nabla_{\theta_n} \log \pi_{\theta_n^t}(a^t | s^t), \\
    &\theta_n^{t+1} = P_{\Theta_n}\left[\theta_n^t - \beta^t A_n^t \psi_n^t\right]. \nonumber
\end{align}
where $\beta^t$ is the actor learning rate. The parameter $\theta_n$ is updated via projected gradient descent, where $P_{\Theta_n}$ projects onto set $\Theta_n$.
In the second phase, agents locally estimate both constraint costs and the Lagrange multipliers. Since global constraint information is unavailable, each agent $n$ maintains local estimates updated  (line 15):
% where $\beta^t$ represents the actor learning rate. The parameter \(\theta_n\) is updated in the descent direction within the set \(\Theta\), where \(P_{\Theta_n}\) denotes the projection onto the set \(\Theta_n\).
% Next, we propose the second phase, which focuses on updating the locally estimated Lagrange multipliers. This phase also involves estimating the constraint cost function. Since agents do not have access to the global constraint cost, these functions are estimated locally using a recursive update as follows:
\begin{equation} 
    \label{dual critic recursion}
    \hat{G}_n^{t+1} = \hat{G}_n^t + \alpha^t \left(g_n^{t+1} - \hat{G}_n^t\right),
\end{equation}
where, \(g^t_n = (g_{n,1}^t, \cdots, g_{n,K}^t)^\top\), \(\hat{G}_n = (\hat{G}_{n,1}, \cdots, \hat{G}_{n,K})^\top\) and \(\hat{G} = (\hat{G}_{1}^\top, \cdots, \hat{G}_{N}^\top)^\top\) are defined for notational simplicity. Finally, we present the update rule for the Lagrange multipliers, which involves taking a step in the ascent direction with respect to $\hat{\mathcal{L}}(\pi_{\theta}, \hat{\lambda})$ for a fixed $\theta$ using the estimated values as follows (lines 20 and 21):
\begin{equation}
    \label{local_lambda_recursion}
    \Tilde{\lambda}_{n}^{t+1} = \sum_{n' \in \mathcal{N}} {\xi}^t_{n,n'} \hat{\lambda}_{n'}^t, \quad \hat{\lambda}_{n}^{t+1} = P_{\Lambda} \left[\Tilde{\lambda}_{n}^t + \gamma^t(\hat{G}_{n}^t - b)\right],
\end{equation}
where, \(\gamma^t\) represents the learning rate for the Lagrange multipliers, and \(P_{\Lambda}\) denotes the projection onto the {%R2.C7
 compact and convex} set \(\Lambda \subseteq \mathbb{R}_+^K\).

 \begin{algorithm}[H]
	\caption{Distributed Primal--Dual Algorithm for CMARL}
	\label{alg:alg1}
	\begin{algorithmic}[1]
		
		\STATE Initialize: $s^0$, $\hat{G}_n^0$, $\hat{\mathcal{L}}_n^0$, $w_n^0$, $\theta_n^0$, $\hat{\lambda}_n^0$ for all $n\in\mathcal{N}$; step sizes $\{\alpha^t\}_{t\ge0}$, $\{\beta^t\}_{t\ge0}$, $\{\gamma^t\}_{t\ge0}$.
		\STATE Each agent executes $a_n^0\sim\pi_{\theta_n^0}(\cdot\mid s^0)$ and observes joint action $a^0=(a_1^0,\ldots,a_N^0)$.
		\STATE Set $t\leftarrow0$.
		
		\REPEAT
		
		\FOR{all $n\in\mathcal{N}$}
		\STATE Observe $s^{t+1}$, scalar cost $c_n^{t+1}$, and constraint cost vector $g_n^{t+1}$.
		\STATE $l_n^{t+1}(\hat{\lambda}_n^t)\leftarrow c_n^{t+1}+(\hat{\lambda}_n^t)^\top(g_n^{t+1}-b)$
		\STATE $\hat{\mathcal{L}}_n^{t+1}\leftarrow \hat{\mathcal{L}}_n^{t}+\alpha^t\!\left(l_n^{t+1}(\hat{\lambda}_n^t)-\hat{\mathcal{L}}_n^{t}\right)$
		\STATE Sample and execute $a_n^{t+1}\sim\pi_{\theta_n^t}(\cdot\mid s^{t+1})$.
		\ENDFOR
		
		\STATE Observe joint action $a^{t+1}=(a_1^{t+1},\ldots,a_N^{t+1})$.
		
		\FOR{all $n\in\mathcal{N}$} 
		\STATE \vspace{0.2em}    \scriptsize  $\delta_n^{t}\xleftarrow[]{}l_n^{t+1}(\hat{\lambda}_n^t)-\hat{\mathcal{L}}_{n}^t+Q_{\lambda}(s^{t+1},a^{t+1};w_n^t)-Q_{\lambda}(s^t,a^t;w_n^t)$ 
		\STATE \normalsize $\tilde{w}_n^t\leftarrow w_n^t+\alpha^t\delta_n^t\nabla_w Q_{\lambda}(s^t,a^t;w_n^t)$
		\STATE $\hat{G}_n^{t+1}\leftarrow \hat{G}_n^{t}+\alpha^t(g_n^{t+1}-\hat{G}_n^{t})$
		\STATE  \scriptsize $A_n^t \xleftarrow{} Q_{\lambda}(s^{t}, a^{t}; w_n^t) - \sum_{a_n \in \mathcal{A}_n} \pi_{\theta_n^t}(a_n | s^{t+1}) Q_{\lambda}(s^t, a^t); w_n^t)$  \vspace{-0.7em}
		\STATE \normalsize $\psi_n^t\leftarrow\nabla_{\theta_n}\log\pi_{\theta_n^t}(a_n^t\mid s^t)$
		\STATE $\theta_n^{t+1}\leftarrow P_{\Theta_n}\!\left[\theta_n^t-\beta^t A_n^t\psi_n^t\right]$
		\STATE $w_n^{t+1}\leftarrow\sum_{n'\in\mathcal{N}}\xi_{n,n'}^t\,\tilde{w}_{n'}^t$
		\STATE $\tilde{\lambda}_n^t\leftarrow\sum_{n'\in\mathcal{N}}\xi_{n,n'}^t\,\hat{\lambda}_{n'}^t$
		\STATE $\hat{\lambda}_n^{t+1}\leftarrow P_\Lambda\!\left[\tilde{\lambda}_n^t+\gamma^t(\hat{G}_n^t-b)\right]$
		\ENDFOR
		
		\STATE $t\leftarrow t+1$
		
		\UNTIL convergence
	\end{algorithmic}
\end{algorithm}

\section{Theoretical Results}
This section consists of two parts. In the first part, we prove the convergence of Algorithm 1 to a stationary point of the update rules in \eqref{critic update}-\eqref{local_lambda_recursion}. In the second part, we analyze the sub-optimality of the convergent point by deriving an upper bound on the gap between the obtained solution and the optimal solution.
\subsection{Convergence of Algorithm}
\begin{assumption}
    \label{ass: step size}
    The learning rates \(\alpha^t\), \(\beta^t\), and \(\gamma^t\) should satisfy the following conditions:
    $$\sum_{t=0}^\infty \alpha^t + \beta^t + \gamma^t = \infty, \quad \sum_{t=0}^\infty (\alpha^t)^2 + (\beta^t)^2 + (\gamma^t)^2 < \infty,$$
    $$\lim_{t \to \infty} \frac{\beta^t}{\alpha^t} = 0, \quad \lim_{t \to \infty} \frac{\gamma^t}{\beta^t} = 0.$$
\end{assumption}

{
%R2.C5
Assumption \ref{ass: step size} is standard to ensure the convergence of stochastic recursions in three different time scales\cite[Assumption 10.6]{Bhatnagar_stochastic}, for updating critic, actor and $\hat{\lambda}_n$.}

\begin{assumption}
   (Slater's condition). There exists some policy $\widetilde{\pi}\in\mathcal{P}$ and constants ${\sigma}_k>0$ such that $G_k(\widetilde{\pi}) \geq$ $b_k+{\sigma}_k$ for all $k\in\mathcal{K}$.
\end{assumption}

{
% R2.C5
This assumption ensures that the interior of the feasible policy set is non-empty, a standard condition in optimization.}
\begin{assumption}
    \label{assumptions on matrix C}
    Consider the sequence of nonnegative random matrices \(\{{\Xi}^t\}_{t \geq 0}\), which satisfies the following conditions:
    \begin{enumerate}
        \item Each matrix \({\Xi}^t\) is doubly stochastic, i.e., \({\Xi}^t \mathbf{1} = \mathbf{1}\) and \(\mathbf{1}^\top{\Xi}^t = \mathbf{1}^\top\).
        \item There exists a constant \(\eta \in (0,1)\) such that for the matrix \({\Xi}^t\), all diagonal elements and any other positive entries satisfy ${\xi}^t_{n,n'} \geq \eta$. \label{decompose Ct}
        \item The spectral norm of the matrix \(\mathbb{E}\left[{{\Xi}^t}^\top \cdot \left(\mathbf{I}_N - \frac{\mathbf{1} \mathbf{1}^\top}{N}\right) \cdot {\Xi}^t\right]\) is denoted by \(\rho\), and it is assumed that \(\rho \in [0,1)\).
        \item Given the \(\sigma\)-algebra generated by the random variables up to time \(t\), the matrix \({\Xi}^t\) is conditionally independent of the variables \(c_n^{t+1}\) and \(g_n^{t+1}\) for all \(n \in \mathcal{N}\).
    \end{enumerate}
\end{assumption}
Assumption \ref{assumptions on matrix C} is a typical condition for ensuring the convergence of networked algorithms \cite{fullydecentralized}. 
\begin{assumption}
\label{actor convergence assumption}
The set \(\Theta\) is sufficiently large such that there exists at least one local minimum of \(\hat{\mathcal{L}}(\pi_{\theta}, \hat{\lambda})\) in its interior, given \(\hat{\lambda} \in \Lambda^N\).
\end{assumption}

This assumption is standard in the analysis of actor-critic algorithms, as it facilitates the convergence analysis of the algorithm. In practice, however, the parameter \(\theta\) can be updated without the need for projection, as discussed in \cite{fullydecentralized}.

\begin{theorem} 
    \label{lambda_perp}
    { %R2C6
    (Consensus of Locally Estimated Lagrange Multipliers). 
    Under Assumptions \ref{non_zero policy}-\ref{actor convergence assumption},}
    the disagreement component of the locally estimated Lagrange multipliers converges to zero almost surely, i.e., \(\lim_{t \to \infty} \hat{\lambda}_\perp^t = 0\) a.s.
\end{theorem}

\textit{Proof: }The recursion for the \(\hat{\lambda}\) vector is given by:
\begin{equation}
    \label{lambda hat recursion}
    \hat{\lambda}^{t+1} = P_{\Lambda^N}\left[\left({\Xi}^t \otimes \mathbf{I}_K\right)\hat{\lambda}^t + \gamma^t(\hat{G}^t - b)\right].
\end{equation}
To analyze the disagreement component, define:
$$D^{t+1} := \frac{P_{\Lambda^N}\left[\left({\Xi}^t \otimes \mathbf{I}_K\right)\hat{\lambda}^t + \gamma^t(\hat{G}^t - b)\right] - \left({\Xi}^t \otimes \mathbf{I}_K\right)\hat{\lambda}^t}{\gamma^t}.$$
The disagreement component \(\hat{\lambda}_{\perp}^{t+1}\) is then given by
$\hat{\lambda}_{\perp}^{t+1} = J_{\perp}\left[\left({\Xi}^t \otimes \mathbf{I}_K\right)\hat{\lambda}^t + \gamma^t D^{t+1}\right].$
Given that \(J_{\perp}(1 \otimes \langle \lambda \rangle) = 0\), we can further simplify to obtain:
\footnotesize$$\hat{\lambda}_{\perp}^{t+1} = \left[\left(\mathbf{I}_N - \frac{\mathbf{1}\mathbf{1}^\top}{N}\right){\Xi}^t \otimes \mathbf{I}_K\right]\hat{\lambda}_{\perp}^t + \gamma^t \left[\left(\mathbf{I}_N - \frac{\mathbf{1}\mathbf{1}^\top}{N}\right) \otimes \mathbf{I}_K\right] D^{t+1}.$$
\normalsize Our goal is to show that \(\mathbb{E}\left[\|(\gamma^{t})^{-1}\hat{\lambda}_\perp^{t}\|^2\right]\) is bounded. Consider the filtration \(\mathcal{F}^t = \sigma\left(\{s^\tau, a^\tau, \hat{\lambda}^\tau, g_n^\tau, \hat{G}_n^\tau, {\Xi}^{\tau-1}\}_{\tau \leq t, n \in \mathcal{N}}\right)\). {Also, It should be noted that the expectations are taken of random variables defined the filtration in this proof.} Then, we have:
\begin{multline}
    \mathbb{E}\left[\|{(\gamma^{t+1})}^{-1}\hat{\lambda}_\perp^{t+1}\|^2 \mid \mathcal{F}^t\right] = \\
    \left(\frac{\gamma^t}{\gamma^{t+1}}\right)^2 \Bigg\{ \mathbb{E} \left[\left({(\gamma^{t})}^{-1}\hat{\lambda}_\perp^{t}\right)^\top \mathbf{X}^t \left({(\gamma^{t})}^{-1}\hat{\lambda}_\perp^{t}\right) \mid \mathcal{F}^t \right] \\
    +2\mathbb{E}\left[\left({(\gamma^{t})}^{-1}\hat{\lambda}_\perp^{t}\right)^\top \mathbf{Y}^t D^{t+1} \mid \mathcal{F}^t \right]  \\
    +\mathbb{E}\left[\left(D^{t+1}\right)^\top \mathbf{Z}^t  D^{t+1} \mid \mathcal{F}^t \right] \Bigg\},
\end{multline}
where \(\mathbf{X}^t = \left({{\Xi}^t}^\top \left(\mathbf{I}_N - \frac{\mathbf{1}\mathbf{1}^\top}{N}\right){\Xi}^t \right) \otimes \mathbf{I}_K\), \(\mathbf{Y}^t = {{\Xi}^t}^\top \left(\mathbf{I}_N - \frac{\mathbf{1}\mathbf{1}^\top}{N}\right) \otimes \mathbf{I}_K\), and \(\mathbf{Z}^t = \left(\mathbf{I}_N - \frac{\mathbf{1}\mathbf{1}^\top}{N}\right) \otimes \mathbf{I}_K\). It is straightforward to show that the maximum eigenvalue of \(\mathbf{Z}^t\) is less than or equal to 1. Utilizing Assumption \ref{assumptions on matrix C}, along with the Cauchy-Schwarz inequality and the Rayleigh quotient inequality, and considering the doubly stochastic property of \({\Xi}^t\), which ensures that \(\|\mathbf{Y}^t\|^2 \leq 1\), we obtain the following inequality:
\begin{multline}
    \label{Exp_inequality}
    \mathbb{E}\left[\|{(\gamma^{t+1})}^{-1}\hat{\lambda}_\perp^{t+1}\|^2 \mid \mathcal{F}^t \right] \leq \left(\frac{\gamma^t}{\gamma^{t+1}}\right)^2 \bigg\{\rho \|{(\gamma^{t})}^{-1}\hat{\lambda}_\perp^{t}\|^2 \\ 
    + 2 \|{(\gamma^{t})}^{-1}\hat{\lambda}_\perp^{t}\| \mathbb{E}\left[\|D^{t+1}\|^2 \mid \mathcal{F}^t\right]^{\frac{1}{2}} + \mathbb{E}\left[\|D^{t+1}\|^2 \mid \mathcal{F}^t\right] \bigg\}.
\end{multline}
Now, we can show that \(\sup_t\mathbb{E}[{\|D^{t+1}\|}^2 | \mathcal{F}^t] < \infty\). At each iteration, \(\hat{\lambda}_n^t\) for \(\forall n \in \mathcal{N}\) is a point in the convex set \(\Lambda\). Additionally, the local consensus update for each agent \(\sum_{n'\in\mathcal{N}}{\xi}^t_{n,n'}\hat{\lambda}_{n}^t\) is a convex combination of \(\hat{\lambda}_n^t\). Consequently, it holds that \(\mathcal{P}_{\Lambda^N}[({\Xi}^t\otimes I)\hat{\lambda}^t] = ({\Xi}^t\otimes I)\hat{\lambda}^t\). Also, from the non-expansive property of convex projection:
\begin{multline*}
\|\mathcal{P}_{\Lambda^N}[({\Xi}^t\otimes I)\hat{\lambda}^t+\gamma^t(\hat{G}^t-b)]-\mathcal{P}_{\Lambda^N}[({\Xi}^t\otimes I)\hat{\lambda}^t]\|\leq \gamma^t \| \hat{G}^t-b \| \\
\Rightarrow  \|D^t\|\leq\|\hat{G}^t-b\|\Rightarrow \|D^t\|^2\leq\|\hat{G}^t\|^2+\|b\|^2.
\end{multline*}
Using Lemma \ref{G_hat is Bounded}, there exists some constant \(M_1\) such that \(\mathbb{E}[{\|D^{t+1}\|}^2 | \mathcal{F}^t]\leq M_1\).
Considering the bound \(M_1\) and taking expectations { of} both sides of \eqref{Exp_inequality}, and applying Jensen's inequality, we have:
\begin{multline}
    \mathbb{E}\left[\|{(\gamma^{t+1})}^{-1}\hat{\lambda}_\perp^{t+1}\|^2\right]\leq \left(\frac{\gamma^t}{\gamma^{t+1}}\right)^2 \bigg(\rho\mathbb{E}\left[\|{(\gamma^{t})}^{-1}\hat{\lambda}_\perp^{t}\|^2 \right] \\
    + 2\sqrt{M_1} \sqrt{\mathbb{E}\left[\|{(\gamma^{t})}^{-1} \hat{\lambda}_\perp^{t}\|^2\right]} + M_1 \bigg).
\end{multline}
Let \(x^{t}=\mathbb{E}\left[\|{(\gamma^{t})}^{-1}\hat{\lambda}_\perp^{t}\|^2 \right]\) for simplicity. Since \(\rho<1\) and \(\lim_{t \to \infty} \left(\frac{\gamma^t}{\gamma^{t+1}}\right)^2 = 1\), there exists a sufficiently large \(t_0\) such that \(\left(\frac{\gamma^t}{\gamma^{t+1}}\right)^2 \rho \leq(1-\epsilon)\) for any \(t \geq t_0\) and for some \(\epsilon>0\). Thus,
\begin{equation}
    \label{xt inequality1}
    x^{t+1}\leq  (1-\epsilon)x^t + 2\sqrt{M_1} \sqrt{x^t} + M_1.
\end{equation}
We define the function \(\psi:x\rightarrow  a\sqrt{x} + b\). With straightforward analysis, it can be shown that for given parameters \(a,b,c\in \mathbb{R}_{{++}}\), there exist \(p,q \in \mathbb{R}_{{++}}\) such that \(\psi(x)\leq cx+ p.\mathds{1}_{\{x<q\}}\), where $\mathds{1}_{\{\cdot\}}$ is an indicator function. Accordingly, using the inequality in \eqref{xt inequality1}, there exist positive constants \(M_2\) and \(m\) that satisfy:
\begin{equation}
    \label{xt_inequality2}
    x^{t+1}\leq(1-\epsilon)x^t + \frac{\epsilon}{2}x^t + m.\mathds{1}_{\{x^t<M_2\}} = (1-\frac{\epsilon}{2})x^t + m.\mathds{1}_{\{x^t<M_2\}}
\end{equation}
for all \(t\geq t_0\). By recursively substituting \eqref{xt_inequality2} { and induction}, it holds that \(x^t\leq(1-\frac{\epsilon}{2})^{t-t_0}x^{t_0} + \frac{2m}{\epsilon}\) for all $t>t_0$. Thus, we can conclude \(\mathbb{E}\left[\|{(\gamma^{t})}^{-1}\hat{\lambda}_\perp^{t}\|^2 \right]\leq M_3\) for a positive constant \(M_3\). Equivalently, \(\mathbb{E}\left[\|\lambda_\perp^{t}\|^2\right]\leq M_3(\gamma^{t})^2\). Summing over \(t\), we obtain $\sum_t \mathbb{E}\left[\|\lambda_\perp^{t}\|^2\right]\leq M_3 \sum_t (\gamma^{t})^2<\infty.$
Finally, by Fubini's theorem, we have \(\mathbb{E}\left[\sum_t \|\lambda_\perp^{t}\|^2\right]<\infty\). Hence, \(\sum_t \|\lambda_\perp^{t}\|^2<\infty\).\QEDA

\begin{proposition} 
{
(Convergence of Distributed Actor-Critic Algorithm). 
\label{actor-critic convergence proposition} 
%R2.C5
 Under Assumptions \ref{non_zero policy}–\ref{actor convergence assumption}, and for {an estimate of the Lagrange multipliers $\hat{\lambda}_n$ generated by \eqref{local_lambda_recursion}}, the distributed actor-critic algorithm employing linear function approximation and local update rules \eqref{critic update}–\eqref{actor_update} guarantees that the policy parameters $\theta_n$, for all agents $n \in \mathcal{N}$, converge almost surely to a point in the set of asymptotically stable equilibrium points of the projected dynamical system defined by}
% Suppose \(w_\theta\) represents the minimizer of the Mean Square Projected Bellman Error (MSPBE) corresponding to the policy \(\pi_\theta\), which will be the consensus point of critics' weights. Under Assumptions \ref{non_zero policy}-\ref{actor convergence assumption}, for a given locally estimated Lagrange multipliers \(\hat{\lambda}\), the distributed actor-critic algorithm with linear critic approximation and local update rules \eqref{critic update}-\eqref{actor_update} ensures that the policy parameters \(\theta_n\) for all \(n \in \mathcal{N}\) converge almost surely to a point within the set of asymptotically stable equilibrium points of the projected dynamical system \eqref{actor dynamics}.
%
%(Original Version) Under Assumptions \ref{non_zero policy}-\ref{actor convergence assumption}, for a given locally estimated Lagrange multipliers \(\hat{\lambda}\), the distributed actor-critic algorithm with linear critic approximation and local update rules \eqref{critic update}-\eqref{actor_update}, when applied to minimize the decomposed Lagrangian objective, ensures that the policy parameters \(\theta_n\) for all \(n \in \mathcal{N}\) converge almost surely to a point within the set of asymptotically stable equilibrium points of the projected dynamical system \eqref{actor dynamics}. Here, \(w_\theta\) represents the minimizer of the Mean Square Projected Bellman Error (MSPBE) corresponding to the policy \(\pi_\theta\). Additionally, the critic weights achieve consensus at the point \(w_\theta\) almost surely.
    \begin{equation}
        \label{actor dynamics}
        \dot{\theta}_n=\hat{P}_{\Theta}\left[-\nabla_{\theta_n} \hat{\mathcal{L}}(\pi_{\theta},\hat{\lambda})-\mathbb{E}^{\pi_{\theta}}\left[w_\theta^\top \phi^t -Q^{\pi_\theta}\left(s^t, a^t\right)\right]\psi^t_n\right], 
    \end{equation}
\end{proposition}
{where $w_\theta$ denotes the unique minimizer of the Mean Square Projected Bellman Error (MSPBE) associated with the policy $\pi_\theta$, representing the consensus value of the critic parameters.}

\textit{Proof:}
Using \eqref{decompose Ct} from Assumption \ref{assumptions on matrix C}, there exists a constant $\bar{\eta} \in (0,1)$ such that
${\Xi}^t = \mathbf{I}_N + \bar{\eta}(\bar{{\Xi}}^t - \mathbf{I}_N)$.
Substituting this into \eqref{local_lambda_recursion}, the update for each $n \in \mathcal{N}$ can be rewritten from the actor timescale perspective as
\begin{multline}
\hat{\lambda}_{n}^{t+1}
=
P_{\Lambda}\bigg[
\hat{\lambda}_n^t
+
\beta^t
\bigg(
\underbrace{
\frac{\gamma^t}{\beta^t}(\hat{G}_{n}^t - b)
+
\frac{\bar{\eta}}{\beta^t}
\sum_{n'\in \mathcal{N}}
\bar{\xi}^t_{n,n'}
(\hat{\lambda}_{n'}^t - \hat{\lambda}_{n}^t)}_{(i)}
\bigg)
\bigg].
\end{multline}

From Theorem \ref{lambda_perp}, the consensus error vanishes almost surely: $\lim_{t \to \infty}
(\hat{\lambda}_{n'}^t - \hat{\lambda}_n^t)
=
0,
\text{a.s., for all } n,n' \in \mathcal{N}$. Therefore, for any $\epsilon > 0$, there exists a sufficiently large time $t'$ such that, almost surely,
$
\sup_{r \ge t'}
\left\|
\sum_{\tau=t'}^{r}
\bar{\eta}
\sum_{n'\in \mathcal{N}}
\bar{\xi}^\tau_{n,n'}
(\hat{\lambda}_{n'}^\tau - \hat{\lambda}_n^\tau)
\right\|
\le \epsilon.
$ Equivalently,
\begin{equation}
\label{constant lambda}
\lim_{t \to \infty}
\mathbb{P}
\left(
\sup_{r \ge t}
\left\|
\sum_{\tau=t}^{r}
\bar{\eta}
\sum_{n'\in \mathcal{N}}
\bar{\xi}^\tau_{n,n'}
(\hat{\lambda}_{n'}^\tau - \hat{\lambda}_n^\tau)
\right\|
\ge \epsilon
\right)
=
0.
\end{equation}
Furthermore, by boundedness of the estimates, we have
$\sup_t \|\hat{G}_n^t - b\| < \infty,$ and by Assumption \ref{ass: step size}, $\lim_{t \to \infty} \frac{\gamma^t}{\beta^t} = 0.$ Hence, both terms of ($i$) in the recursion vanish asymptotically. 

Therefore, the recursion satisfies the conditions of the Kushner–Clark lemma (Theorem 5.3.1 in \cite{kushner1978stochastic}), and the interpolated trajectory of $\hat{\lambda}_n^t$ tracks the ordinary differential equation
$\frac{d\hat{\lambda}_n}{dt} = 0$ with rate $\beta^t$, i.e. with the same timescale as the actor update in \eqref{actor_update}. This means that due to the condition $\gamma^t/\beta^t \to 0$, the update of multiplier $\hat{\lambda}_n^t$  operates on a strictly slower timescale than the actor update. Consequently, $\hat{\lambda}_n^t$ can be treated as quasi-static (constant) from the perspective of the actor dynamics.

Now, for a constant multiplier $\hat{\lambda}$, the estimated Lagrangian
$\hat{\mathcal{L}}(\pi_{\theta}, \hat{\lambda})$
reduces to an unconstrained multi-agent reinforcement learning objective.
Therefore, by Theorems 4.6 and 4.7 of \cite{fullydecentralized}, 
{which rely on $\beta^t/\alpha^t \to 0$ to ensure convergence via a two-timescale argument,}
 $\theta_n$ converges almost surely to a point in the set of asymptotically stable equilibria of \eqref{actor dynamics}. \QEDA

\begin{remark}
\label{remark2}
    According to Proposition \ref{actor-critic convergence proposition}, for a good choice of the basis function of the critic approximator, which satisfies $\sup_{\theta} \mathbb{E}^{\pi_{\theta}}\left[{w_\theta}^\top \phi^t  - Q^{\pi_\theta}\left(s^t, a^t\right)\right] < \epsilon$, for a sufficiently small $\epsilon$, the actor converges to the $\epsilon$-neighborhood of the local minima of $\hat{\mathcal{L}}(\theta,\hat{\lambda})$.
\end{remark}

\begin{assumption}
{
    \label{theta_bar function} There exists a unique and continuous function \(\bar{\theta}(\hat{\lambda}): \Lambda^{N} \rightarrow \mathbb{R}^{\dim(\theta)}\) that maps each multiplier estimate $\hat{\lambda}$ to the fixed point of the actor dynamics defined in \eqref{actor dynamics}.}
\end{assumption}
 
\begin{remark}
\label{remark3}
    According to Proposition \ref{actor-critic convergence proposition} and Remark \ref{remark2}, the Jacobian $\nabla_{\theta} \hat{\mathcal{L}}$ is positive definite at 
    { %R2.C4
    a nearby neighborhood} of the local minima of $\hat{\mathcal{L}}(\theta,\hat{\lambda})$ and hence, it can be concluded from the implicit function theorem that Assumption \ref{theta_bar function} is valid.
\end{remark}

{
%R2.C5
Proposition \ref{actor-critic convergence proposition} establishes the convergence of the actor-critic algorithm for a fixed $\hat{\lambda}$. Subsequently, Lemma \ref{G_hat is Bounded} and Theorem \ref{lambda_perp} show that $\hat{\lambda}$ reaches consensus across agents.
}

\begin{lemma}
    \label{G_hat is Bounded}
    Under Assumptions \ref{non_zero policy} and \ref{signal boundedness}, it holds that $\sup_t \|\hat{G}^t\|^2 < \infty$ almost surely for all $n \in \mathcal{N}$.
\end{lemma}
\textit{Proof: } The recursion \eqref{dual critic recursion} is driven only by immediate local costs, and these local costs are independent of each other's constraints. By applying Lemma 5.2 from \cite{fullydecentralized}, which addresses the boundedness of updates for local $\hat{G}^t_{n,k}$ independently, we can conclude that $\sup_t \hat{G}^t_{n,k} < \infty$ almost surely. Consequently, $\sup_t \|\hat{G}^t\|<\infty$, and therefore $\sup_t \|\hat{G}^t\|^2 < \infty$. \QEDA

% \begin{assumption}  
%     Suppose the total number of policy parameters is \(T\). There exists a unique continuous function \(\bar{\theta}(\hat{\lambda}): \Lambda^{N} \rightarrow \mathbb{R}^T\), which represents the converged point of the actor dynamics in \eqref{actor dynamics} for a given \(\hat{\lambda}\).
% \end{assumption}

\begin{lemma}
    \label{dual critic convergence}
    (Proposition 4.2 \cite{bhatnagar2012online}) Under Assumption \ref{signal boundedness}, for a fixed policy $\theta$, every dual critic recursion \eqref{dual critic recursion} for all $n \in \mathcal{N}$ converges almost surely to
    $\bar{G}_{n}(\pi_{\theta}) := \lim_{T \to \infty} \frac{1}{T}\mathbb{E}^{\pi_{\theta}}\left[\sum_{t=0}^{T} g_n^{t+1}\right]\textit{.}$
\end{lemma}
Let the projected dynamical system for \( \hat{\lambda} \) be defined as follows:
\begin{equation}
    \label{lambda dynamic}
    \frac{d \hat{\lambda}}{dt} = \hat{P}_{\Lambda^N}\left[\bar{G}(\pi_{\bar{\theta}(\hat{\lambda})}) -\mathbf{1} \otimes b\right],
\end{equation}
where \( \bar{G}(\pi_{\theta}) = \left(\bar{G}_1(\pi_{\theta})^\top, \dots, \bar{G}_N(\pi_{\theta})^\top\right)^\top \).

We define the set $F \subseteq \Lambda^N$ as the compact set of all asymptotically stable equilibrium points of the projected dynamical system given in \eqref{lambda dynamic}.
Building on the definition of the set  $F$, Theorem \ref{convergence of Lagrangian multipliers} shows that the consensus vector of the locally estimated Lagrange multipliers converges.

\begin{theorem}
\label{convergence of Lagrangian multipliers}
Under Assumptions \ref{ass: step size}, \ref{assumptions on matrix C} and \ref{theta_bar function}, it is established that $\lim_{t\to\infty} \langle \lambda^t \rangle = \bar{\lambda}$ almost surely, where $\mathbf{1} \otimes \bar{\lambda} \in F$.
\end{theorem}
\textit{Proof}: We rewrite \eqref{local_lambda_recursion} for each $n \in \mathcal{N}$ by adding and subtracting $\bar{G}_n(\pi_{\theta})$:
\begin{multline}
    \label{single_agent_lambda_recurssion}
    \hat{\lambda}_{n}^{t+1} = P_{\Lambda}\bigg[\hat{\lambda}_n^t + \gamma^t\bigg((\bar{G}_n(\pi_{\theta})-b) + \frac{\bar{\eta}}{\gamma^t} \sum_{n'\in \mathcal{N}} \bar{{\xi}}^t_{n,n'}(\hat{\lambda}_{n'}^t - \hat{\lambda}_{n}^t) \\
    + (\hat{G}_{n}^t - \bar{G}_n(\pi_{\theta}))\bigg)\bigg].
\end{multline}
By Theorem \ref{lambda_perp}, \eqref{constant lambda} holds as well. Furthermore, by Lemma \ref{dual critic convergence}, $\lim_{t\to\infty} (\hat{G}_{n}^t - \bar{G}_n(\pi_{\theta})) = 0$ almost surely. Therefore, by applying Kushner-Clark Lemma under Assumptions \ref{ass: step size} and \ref{theta_bar function}, the recursion \eqref{local_lambda_recursion} converges almost surely to a point in the set $F$. Consequently, as per Theorem \ref{lambda_perp}, we conclude that $\lim_{t\to\infty} \hat{\lambda}^t = \mathbf{1} \otimes \bar{\lambda}$ almost surely.\QEDA

% By Theorem \ref{lambda_perp}, we have $\lim_{t\to\infty}(\hat{\lambda}_{n'}^t - \hat{\lambda}_n^t) = 0$ almost surely, for all $n, n' \in \mathcal{N}$. Consequently, for any $\epsilon > 0$, there exists a sufficiently large $t'$ such that $\sup_{r \geq t'} \left\|\sum_{\tau=t'}^r \bar{\eta}\sum_{n' \in \mathcal{N}} \bar{{\xi}}^\tau_{n,n'}(\hat{\lambda}_{n'}^\tau - \hat{\lambda}_{n}^\tau)\right\| \leq \epsilon$ almost surely. Equivalently,
% \begin{equation}
%     \lim_{t\to\infty} \mathbb{P}\left(\sup_{r \geq t} \left\|\sum_{\tau=t}^r \bar{\eta} \sum_{n' \in \mathcal{N}} \bar{{\xi}}^\tau_{n,n'}(\hat{\lambda}_{n'}^\tau - \hat{\lambda}_{n}^\tau)\right\| \geq \epsilon\right) = 0.
% \end{equation}

The following proposition introduces a condition under which the feasibility of the constraints is met at the consensus value of 
Lagrange multipliers, i.e., $\bar \lambda$.

\begin{proposition}
    Suppose \(\bar{\lambda} \in \text{Int}(\Lambda)\). Then all constraints are satisfied.
\end{proposition}

\textit{Proof:} This can be proven by contradiction. Assume the claim is not true. Then, for a sufficiently small $\varepsilon$ and for all \(n \in \mathcal{N}\), we have $P_{\Lambda}\left[\bar{\lambda} + \varepsilon \left({G}_n(\pi_{\theta(\bar{\lambda})}) - b\right)\right] = \bar{\lambda} + \varepsilon \left({G}_n(\pi_{\theta(\bar{\lambda})}) - b\right).$
Using the definition of projected dynamical system for the equation \eqref{lambda dynamic}, we obtain:
\begin{align}
    \hat{P}_{\Lambda^N}\bigg[\bar{G}&(\pi_{\theta(\mathbf{1} \otimes \bar{\lambda})}) - \mathbf{1} \otimes b\bigg] \nonumber\\
    &= \lim_{ \varepsilon \rightarrow 0^+} \left[\frac{\mathbf{1} \otimes \bar{\lambda} + \varepsilon(\bar{G}(\pi_{\theta(\mathbf{1} \otimes \bar{\lambda})}) -\mathbf{1} \otimes b) - \mathbf{1} \otimes \bar{\lambda}}{\varepsilon}\right] \nonumber \\
    &= \bar{G}(\pi_{\theta(\mathbf{1} \otimes \bar{\lambda})}) -\mathbf{1} \otimes b > 0.
\end{align}

This leads to a contradiction since \(\mathbf{1} \otimes \bar{\lambda}\) is an interior equilibrium point for the dynamic \eqref{lambda dynamic}. Hence, the original claim is true. \QEDA

\subsection{Duality Gap Evaluation at Equilibrium Point}

According to \cite{constrainedMARLHardness}, CMARL problems without parameterization can exhibit a strictly positive duality gap $\bar{\Delta} := \mathbf{P}^* - \mathbf{D}^*$, where $\mathbf{P}^* = \inf_{\pi \in \mathcal{P}} \sup_{\lambda \in \mathbb{R}_{+}^K} \mathcal{L}(\pi, \lambda)$ and $\mathbf{D}^* = \sup_{\lambda \in \mathbb{R}_{+}^K} \inf_{\pi \in \mathcal{P}} \mathcal{L}(\pi, \lambda)$. This gap persists in parameterized problems since 
parameterized policy class which is
a subset of the policy space
 $\mathcal{P}$ forms a nonlinear optimization problem.
We analyze errors from individual and product policy parameterizations, and how local convergence of Lagrange multipliers affects the primal-dual discrepancy. The parameterized duality gap is defined as $\bar{\Delta}_{\text{param}} := \boldsymbol{\mathrm{P}}^* - \mathcal{L}(\pi_{\bar{\theta}}, \bar{\lambda})$, where $(\bar{\theta}, \bar{\lambda})$ are the algorithm's convergence points.

The following lemma bounds the optimal Lagrange multipliers $\lambda_k^*$ of the non-parameterized dual problem:

\begin{lemma}
    \label{Lambda}
    (Lemma 1 \cite{constrainedMARLHardness}) Any optimal Lagrange multiplier $\lambda_k^* \in \arg \max_{\lambda \in \mathbb{R}_{+}^K} \inf_{\pi\in\mathcal{P}} \mathcal{L}(\pi, \lambda)$ satisfies the range $\lambda_k^* \leq \lambda_{k}^{\max}:=\frac{1+\bar{\Delta}}{\delta_k}$ for all $k\in\mathcal{K}$.
\end{lemma}

%According to Proposition \ref{Lambda}, we define the set \(\Lambda = [0, \lambda_1^{\max}] \times \cdots \times [0, \lambda_K^{\max}]\), which serves as the domain in which the Lagrange multiplier variables are confined when solving the dual problem. These bounds, or any other estimated bounds, can be applied within our proposed algorithm.

\begin{definition}
    A parameterization \(\pi_{\theta_n}\) is an \(\epsilon_n\)-individual parameterization of policies in \(\mathcal{P}_n\) if, for some \(\epsilon > 0\), there exists a parameter \(\theta_n \in \Theta_n\) for every policy \(\pi_n \in \mathcal{P}_n\) such that
    $\max_{s \in \mathcal{S}} \sum_{a \in \mathcal{A}} \left|\pi_n(a|s) - \pi_{\theta_n}(a|s)\right| \leq \epsilon_n.$
\end{definition}
\begin{definition}
    A parameterization $\pi_\theta$ is an $\epsilon$-product parameterization of policies in $\mathcal{P}$ if, for some $\epsilon>0$, there exists for any $\pi \in \mathcal{P}$ a parameter $\theta \in \Theta$ such that $\max_{s \in \mathcal{S}} \sum_{a\in\mathcal{A}}\left|\pi(a|s)-\pi_\theta(a|s)\right| \leq \epsilon$.
\end{definition}

{
%R2.C5
The following lemma characterizes how the local policy parameterization errors  relates to the global policy parameterization error. 
}

\begin{lemma}
    Assuming \(\epsilon_n\)-individual parameterization for each agent \(n \in \mathcal{N}\), then \(\sum_{n \in \mathcal{N}} \epsilon_n\) is an upper bound for \(\epsilon\) in the \(\epsilon\)-product parameterization policy.
\end{lemma}
\textit{Proof:} By considering the difference between the policy \(\pi(a|s)\) and the parameterized policy \(\pi_\theta(a | s)\)\small:
\begin{multline}
    \max_{s \in \mathcal{S}} \sum_{a \in \mathcal{A}} \left|\pi(a | s) - \pi_\theta(a | s)\right| = \max_{s \in \mathcal{S}} \sum_{a \in \mathcal{A}} \bigg|\pi_{\theta_{-n}}(a_{-n} | s)\bigg(\pi_n(a_n | s)  \\
    - \pi_{\theta_n}(a_n | s)\bigg) + \pi_n(a_n | s)\bigg(\pi_{-n}(a_{-n} | s) - \pi_{\theta_{-n}}(a_{-n} | s)\bigg)\bigg|.
\end{multline}
\normalsize Using the triangle inequality and { applying the point-wise maximum to the right- and left-hand sides of the inequality respectively}, we can bound the above expression by:
\begin{multline}
    \underbrace{\max_{s \in \mathcal{S}} \sum_{a \in \mathcal{A}} \left|\pi_{\theta_{-n}}(a_{-n} | s)\left(\pi_n(a_n | s) - \pi_{\theta_n}(a_n | s)\right)\right|}_{(i)} \\
    + \underbrace{\max_{s \in \mathcal{S}} \sum_{a \in \mathcal{A}} \left|\pi_n(a_n | s)\left(\pi_{-n}(a_{-n} | s) - \pi_{\theta_{-n}}(a_{-n} | s)\right)\right|}_{(ii)}.
\end{multline}
% Let denote the two terms separately for clarity:
% \begin{align*}
%     &(i): \max_{s \in \mathcal{S}} \sum_{a \in \mathcal{A}} \left|\pi_{\theta_{-n}}(a_{-n} | s)\left(\pi_n(a_n | s) - \pi_{\theta_n}(a_n | s)\right)\right|,\\
%     &(ii): \max_{s \in \mathcal{S}} \sum_{a \in \mathcal{A}} \left|\pi_n(a_n | s)\left(\pi_{-n}(a_{-n} | s) - \pi_{\theta_{-n}}(a_{-n} | s)\right)\right|.
% \end{align*}
{
%R2.C10
For the term \((i)\), we have that \((i) \leq \max_{s \in \mathcal{S}} \sum_{a_n \in \mathcal{A}_n} \left|\pi_n(a_n | s) - \pi_{\theta_n}(a_n | s)\right| \leq \epsilon_n\) and For the term \((ii)\), it follows that $(ii)\leq \max_{s \in \mathcal{S}} \sum_{a_{-n} \in \mathcal{A}_{-n}} \left|\pi_{-n}(a_{-n} | s) - \pi_{\theta_{-n}}(a_{-n} | s)\right|.$\\
Since \(\pi_{\theta_{-n}}(a_{-n} | s)\) is an \(\epsilon\)-product parameterized policy for any product policy in \(\Pi_{-n}\), the inequality for term $(ii)$ can be bounded via induction as follows:}
$\max_{s \in \mathcal{S}} \sum_{a_{-n} \in \mathcal{A}_{-n}} \left|\pi_{-n}(a_{-n} | s) - \pi_{\theta_{-n}}(a_{-n} | s)\right| \leq \sum_{n' \in \mathcal{N} \setminus \{n\}} \epsilon_{n'}.$ Combining these results, we obtain $\max_{s \in \mathcal{S}} \sum_{a \in \mathcal{A}} \left|\pi(a | s) - \pi_\theta(a | s)\right| \leq \sum_{n \in \mathcal{N}} \epsilon_n$. \QEDA
\\
{
%R2.C5

Lemmas \ref{d&p} and \ref{L1 distance of two stationary distributions} jointly establish an upper bound on the discrepancy between the stationary distributions of a policy and its parameterized version.}
\begin{lemma}
    \label{d&p}
    (Lemma 3 \cite{average_reward}): 
    The \(L_1\) distance between the stationary distributions \(d^\pi(s)\) and \(d^{\pi'}(s)\), which are induced by the policies \(\pi, \pi' \in \mathcal{P}\), can be upper bounded by:
    \begin{multline}
        D_{L_1}(d^{\pi}(s),d^{\pi'}(s)):=\sum_{s \in \mathcal{S}} \left|d^{\pi}(s) - d^{\pi'}(s)\right| \\
        \leq \left(\kappa^{\star} - 1\right) \max_{s \in \mathcal{S}} \sum_{a \in \mathcal{A}} \left|\pi(a \mid s) - \pi'(a \mid s)\right|,
    \end{multline}
    where \(\kappa^{\star} = \max_{\pi} \kappa^\pi\) and \(\kappa^\pi\) represents the average number of steps required to reach a target state in the Markov chain induced by the policy \(\pi\). 
\end{lemma} 
%In Lemma \ref{d&p}, \(\kappa^\pi\), known as Kemeny’s constant, represents the average number of steps required to reach a target state in the Markov chain induced by the policy \(\pi\). It is calculated as the mean first passage time for each state, weighted according to the steady-state distribution. This constant is also related to the mixing time of the Markov chain, providing a measure of how quickly the chain converges to its stationary distribution.
\begin{lemma}
    \label{L1 distance of two stationary distributions}
    Let \(d(s, a)\) and \(d^\theta(s, a)\) be the occupation measure induced by any stationary policy $\pi$ and \(\pi_\theta \textit{ for } \theta\in\Theta\), respectively, where \(\pi_\theta\) is an \(\epsilon\)-product approximation of \(\pi\). Then, it follows that:
    \begin{equation}
        D_{L_1}(d(s,a),d^{\theta}(s,a))=\sum_{s, a} \left|d(s, a) - d^\theta(s, a)\right| \leq \kappa^{\star}\epsilon.
    \end{equation}
\end{lemma}
\textit{Proof:} Expressing the total variation distance between the stationary distributions \(d(s, a)\) and \(d^\theta(s, a)\):
\begin{equation}
    \sum_{s, a} \left|d(s, a) - d^\theta(s, a)\right| = \sum_{s, a} \left|d(s)\pi(a | s) - d^\theta(s)\pi_\theta(a | s)\right|.
\end{equation}

By adding and subtracting the term $d(s)\pi_\theta(a | s)$ and using triangle inequality we obtain:
%\begin{multline*}
 %   \sum_{s, a} \left|d(s, a) - d^\theta(s, a)\right| = \sum_{s, a} \bigg|\left( d(s)\pi(a | s) - d(s)\pi_\theta(a | s) \right) \\
  %  + \left( d(s)\pi_\theta(a | s) - d^\theta(s)\pi_\theta(a | s)\right)\bigg|.
%\end{multline*}
%Using the triangle inequality, we obtain:
\begin{multline}
    \sum_{s, a} \left|d(s, a) - d^\theta(s, a)\right| \leq \underbrace{\sum_{s, a} d(s)\left|\pi(a | s) - \pi_\theta(a | s)\right|}_{(i')} \\
    + \underbrace{\sum_{s, a} \pi_\theta(a | s)\left|d(s) - d^\theta(s)\right|}_{(ii')}.
\end{multline}
%By evaluate the two terms \((i)\) and \((ii)\):

%: \quad \sum_{s, a} d(s)\left|\pi(a | s) - \pi_\theta(a | s)\right|
{
%R2.C10
Then have
}
$(i') \leq \epsilon \sum_{s} d(s) = \epsilon$, and from Lemma \ref{d&p}, $(ii')= \sum_{s} \left|d(s) - d^\theta(s)\right|   \leq\left(\kappa^{\star} - 1\right) \epsilon.$
%\end{multline*}
Finally
$\sum_{s, a} \left|d(s, a) - d^\theta(s, a)\right| \leq \epsilon + \left(\kappa^{\star} - 1\right) \epsilon = \kappa^{\star} \epsilon.$ \QEDA

The following Lemma provides a bound for the difference in the values of the Lagrangian functions with respect to the \(L_1\) norm of the Lagrangian multipliers and the \(L_1\) distance between the stationary distributions of the induced policies.
\begin{lemma}
    \label{lemma: Lagrangian functions  differnce}
    The difference in the values of the Lagrangian functions \(\mathcal{L}(\pi, \lambda)\) and \(\mathcal{L}(\pi', \lambda')\) for two pairs \((\pi, \lambda)\) and \((\pi', \lambda')\) is bounded by:
    \small
    \begin{equation}
    \mathcal{L}(\pi, \lambda) - \mathcal{L}(\pi', \lambda') \leq \|\lambda - \lambda'\|_1 + \left(1 + K\|\lambda'\|_1\right) D_{L_1}(d^\pi(s, a), d^{\pi'}(s, a)),
    \end{equation}
\end{lemma}
\vspace{1em}
\normalsize
\textit{Proof: }We can express the difference in the Lagrangian values as the sum of two terms:\small
\begin{equation}
    \mathcal{L}(\pi, \lambda) - \mathcal{L}(\pi', \lambda') = \underbrace{\left(\mathcal{L}(\pi, \lambda) - \mathcal{L}(\pi, \lambda')\right)}_{(i'')} + \underbrace{\left(\mathcal{L}(\pi, \lambda') - \mathcal{L}(\pi', \lambda')\right)}_{(ii'')}.
\end{equation}
\normalsize
Then have $(i'') = \sum_{s,a} \left(\sum_{k \in \mathcal{K}} (\lambda_k - \lambda'_k) g_k(s,a)\right) d^{\pi}(s,a) \leq \sum_{k \in \mathcal{K}} |\lambda_k - \lambda'_k| = \|\lambda - \lambda'\|_1.$ Also, $(ii'') = \sum_{s,a} \left(c(s,a) + \sum_{k \in \mathcal{K}} \lambda'_k g_k(s,a)\right) \bigg(d^{\pi}(s,a)- d^{\pi'}(s,a)\bigg) \leq \left(1 + K\|\lambda'\|_1\right) D_{L_1}(d^\pi(s,a), d^{\pi'}(s,a))$. Combining the bounds for \((i'')\) and \((ii'')\), we obtain the desired result. \QEDA

Let the pair \((\pi^*, \lambda^*)\) be the solution to the problem \(\max_{\lambda \in \Lambda} \min_{\pi \in \mathcal{P}} \mathcal{L}(\pi, \lambda)\), and let \(\theta^*\) be the solution to the problem \(\max_{\lambda \in \Lambda} \min_{\theta \in \Theta} \mathcal{L}(\pi_\theta, \lambda)\).

\begin{proposition}
The duality gap for the parameterized problem is bounded as follows:
\begin{equation}
    \bar{\Delta}_{\text{param}} \leq \|\lambda^{\max}\|_1 + (1 + K \|\lambda^{\max}\|_1)\left(\kappa^* \sum_{n \in \mathcal{N}} \epsilon_n + \bar{\epsilon} \right) + \bar{\Delta},
    \end{equation}
where \(\bar{\epsilon} = D_{L_1}(d^{\pi_{\theta^*}}(s,a), d^{\pi_{\bar{\theta}}}(s,a))\), and \(\lambda^{\max} = (\lambda_1^{\max}, \cdots, \lambda_K^{\max})^\top\).
\end{proposition}
\textit{Proof:} Using Lemma \ref{lemma: Lagrangian functions  differnce}, we have\small:
$$\mathcal{L}(\pi^*, \lambda^*) - \mathcal{L}(\pi_{\bar{\theta}}, \bar{\lambda}) \leq \|\lambda^* - \bar{\lambda}\|_1 + (1 + K\|\bar{\lambda}\|_1) D_{L_1}(d^{\pi^*}(s,a), d^{\pi_{\bar{\theta}}}(s,a)).$$
\normalsize 
Since \(D_{L_1}\) is a distance metric, we can apply the triangle inequality
$D_{L_1}(d^{\pi^*}(s,a), d^{\pi_{\bar{\theta}}}(s,a)) \leq D_{L_1}(d^{\pi^*}(s,a), d^{\pi_{\theta^*}}(s,a)) + D_{L_1}(d^{\pi_{\theta^*}}(s,a), d^{\pi_{\bar{\theta}}}(s,a))$ where \(\pi_{\theta^*}\) is an \(\epsilon\)-product parameterization of \(\pi^*\). Finally, using Proposition \ref{Lambda}, we have \(\|\lambda^* - \bar{\lambda}\|_1 \leq \|\lambda^{\max}\|_1\) and \(\|\bar{\lambda}\|_1 \leq \|\lambda^{\max}\|_1\). Substituting these bounds into the earlier inequality and using the relationship between the duality gap and the optimal values, \(\boldsymbol{\mathrm{D}}^* = \boldsymbol{\mathrm{P}}^* - \bar{\Delta}\), we conclude the proof. \QEDA

\section{Experiments}
%The Cournot game is a fundamental economic model that describes an oligopoly market where firms compete by simultaneously selecting their output quantities. Each firm makes its decision based on the expected output of the other firms, aiming to maximize its profit given the production levels of its competitors.

We define a cooperative, stochastic Cournot game to evaluate the algorithm in a complex economic setting. Each agent's cost is given by $c_n(s, a) = -\big(x(s, a)a_n - h a_n\big)$, where $a_n \in [0, 1]$ is the production level, $h = 1$ is the unit price, and the market price $x(s, a) = N - s \sum_{n \in \mathcal{N}} a_n$ depends on the state $s \in (0, 1)$ and agents' actions. This model captures time-varying demand, offering greater realism.

In the above formulation, the original continuous Cournot model is discretized: states and actions are sampled uniformly (10 states from $[0.1, 0.9]$, 10 actions per agent from $[0, 1]$), yielding $10 \times 10^5$ state-action pairs. State transitions follow a binomial distribution, where higher total actions shift probability mass toward lower states, increasing stochasticity.

Constraints are incorporated via agent-specific price bounds $g_n(s, a) = m_n\big(N - s \sum_{n \in \mathcal{N}} a_n\big)$, with weights $m_1 = 0.1$, $m_2 = 0.3$, $m_3 = 0.5$, $m_4 = 0.1$, $m_5 = 0$, and a global bound $b = 0.75$.
Agents use linear critics (20-dimensional parameters, $[0, 1]$-sampled shared features) and softmax policies with linear parameterization (10-dimensional parameters, agent-specific $[0, 1]$-sampled features). Learning rates follow $\alpha^t = t^{-0.6}$ (critic), $\beta^t = t^{-0.75}$ (actor), and $\gamma^t = t^{-0.9}$ (Lagrange multipliers).

The experiment results are depicted in Figures \ref{fig1: Lagrangian multipliers} and \ref{fig2: Costs}. In Figure \ref{fig1: Lagrangian multipliers}, it is shown that the locally estimated Lagrange multipliers reach consensus and subsequently converge, as demonstrated by Theorems \ref{lambda_perp} and \ref{convergence of Lagrangian multipliers}. Additionally, Figure \ref{fig2: Costs} illustrates the global objective cost \(J\) and the global constraint cost \(\hat{G} - b\) during training. This plot demonstrates how effectively the algorithm reduces the objective cost while maintaining constraint violations near zero.
\begin{figure}[h]
    \centering
    \includegraphics[width=0.40\textwidth, height=0.3\textheight, keepaspectratio]{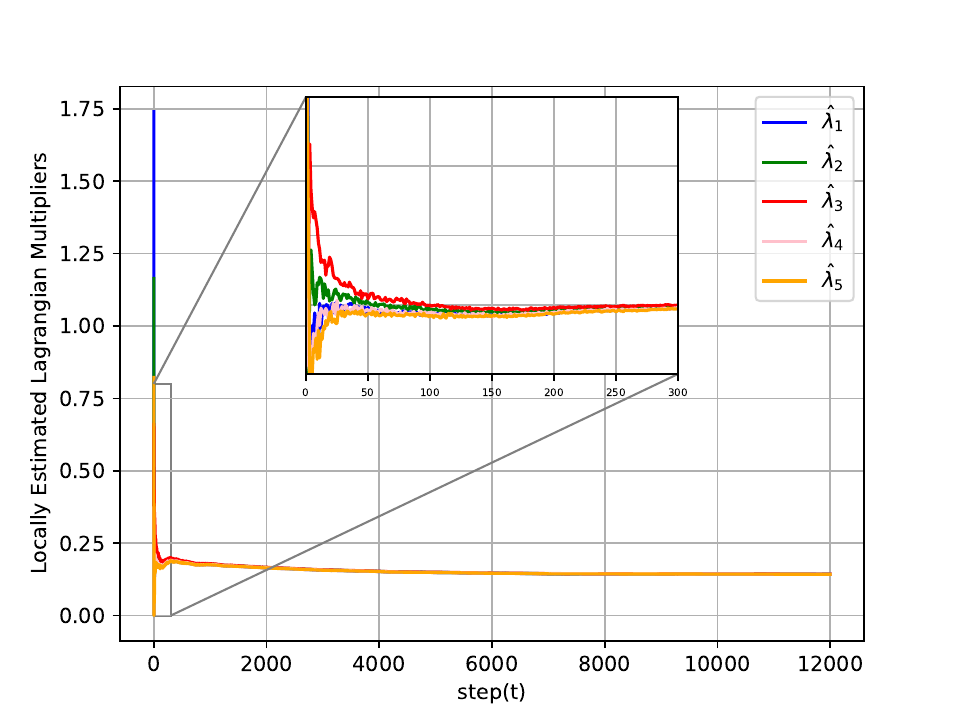}
    \caption{Convergence of locally estimated Lagrangian multipliers during training.}
    \label{fig1: Lagrangian multipliers}
\end{figure}
\begin{figure}[h]
    \centering
    \includegraphics[width=0.45\textwidth, keepaspectratio]{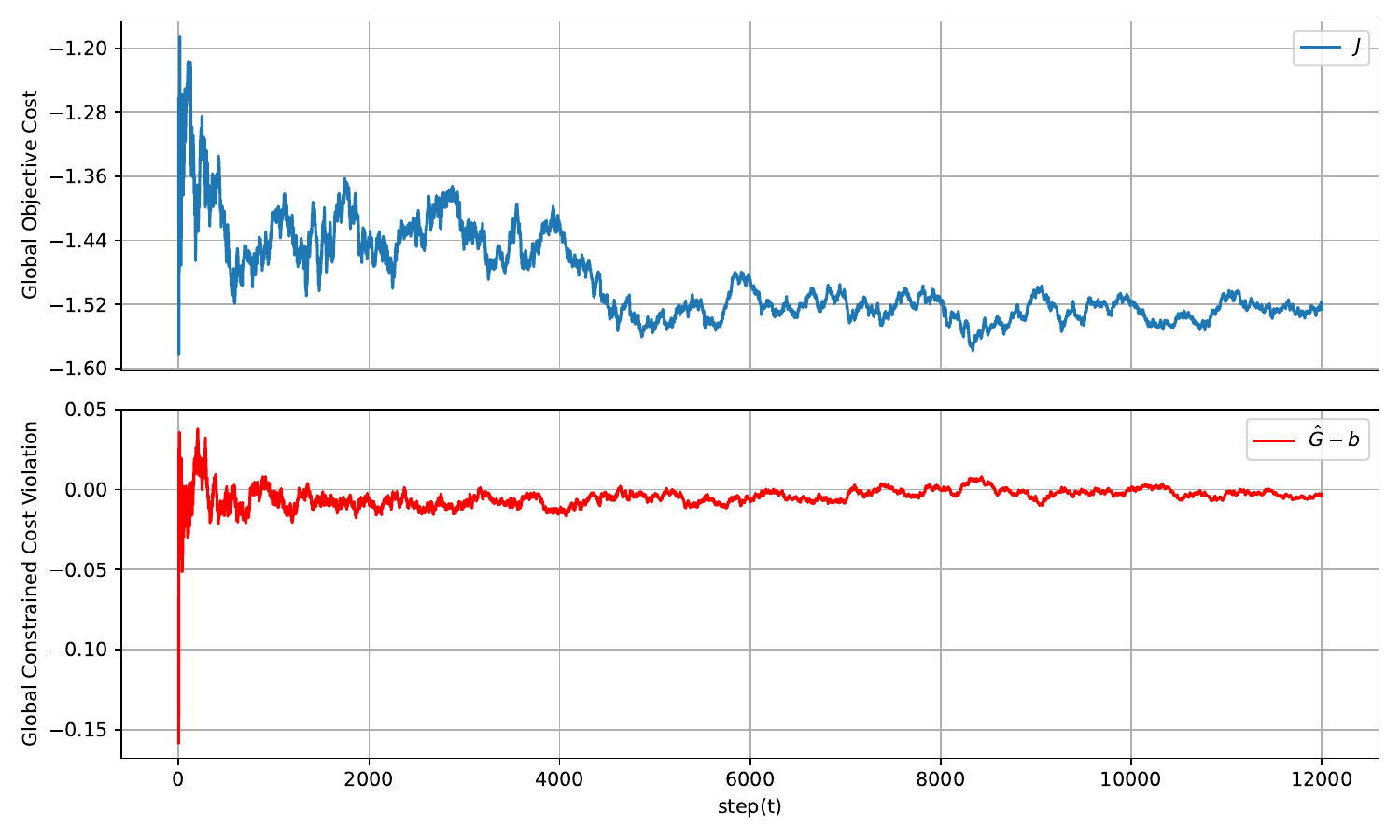}
    \caption{Global objective cost (\(J\)) and global constraint cost (\(\hat{G} - b\)) during training. The plot illustrates how the algorithm seeks to decrease the objective cost while maintaining constraint violations near zero.}
    \label{fig2: Costs}
\end{figure}

\section{Conclusion} 
    This paper introduced a distributed approach for solving cooperative CMARL problems, allowing agents to minimize a global objective function while respecting shared constraints. By leveraging a decentralized primal-dual algorithm based on the actor-critic method, our approach enables each agent to estimate Lagrangian multipliers locally, achieving consensus and ensuring convergence to an equilibrium. We validated the method through a constrained cooperative Cournot game with stochastic dynamics. The results highlight the potential of our method for scalable, decentralized solutions in real-world applications, such as smart grids and autonomous systems. Future work will explore adaptations for more complex environments and dynamic constraints.
 \bibliographystyle{IEEEtran}
% argument is your BibTeX string definitions and bibliography database(s)
%\bibliography{IEEEabrv,../bib/paper}
%
\bibliography{ref}

\end{document}